\documentclass[twocolumn,times,trackchanges, dvipsnames]{aastex631}
\usepackage[utf8]{inputenc}
\usepackage{amsmath}

\defcitealias{Carvalho_FUVFUori_2024ApJ}{C24}

\begin{document}
\title{The FUor Mass Distribution Matches the Solar Neighborhood IMF: Evidence for a Universal Eruptive Phase}

\author{Adolfo S. Carvalho}
\affiliation{Department of Astronomy; California Institute of Technology; Pasadena, CA 91125, USA}
\affiliation{Center for Astrophysics | Harvard \& Smithsonian, Cambridge, MA 02138, USA}
\author{Lynne A. Hillenbrand}
\affiliation{Department of Astronomy; California Institute of Technology; Pasadena, CA 91125, USA}

\begin{abstract}
    Eruptive accretion events are expected to play an important role in the mass buildup stage of individual star formation. FU Ori objects (FUors) experience the most extreme eruptive outbursts, which raise the accretion rate of the disk from $10^{-9}-10^{-8} \ M_\odot \ \mathrm{yr}^{-1}$ to $10^{-5}-10^{-4} \ M_\odot \ \mathrm{yr}^{-1}$ and last for decades. During an outburst, the disk is approximately 100 times brighter than the star, making direct study of the central star impossible. However, the disk is expected to be in Keplerian rotation around the star, enabling indirect constraints on properties of the central source via observations of the disk. Using $1-2.4 \ \mu$m high resolution spectra of several tens of FUors, we demonstrate the expected Keplerian rotation in their inner disks. We then adopt a Keplerian rotational broadening profile to model the line profiles of spectral lines, and focussing on the H-band region, we infer the mass distribution of FUors. We finally show that this mass distribution is consistent with inferred Solar neighborhood initial mass functions, suggesting all young stars undergo a period of FUor outbursts in their pre main-sequence evolution. 
\end{abstract}

%    \lahcomm{regarding the title trial-edit, i strangely do not mind at all the lower case spelling of this,    that bo indicated is the original.  it is similar to the way we would say qso or agn.    among Fuor, FUor, FUor, FUori, and FUs, i kind of like it best.  lee hartmann sent me a message this morning in which he used "Fuor" in the middle of a sentence.....    but put it back the way you like the spelling; i merely wanted to stare at "fuor" in .latex.}

\section{Introduction}\label{sec:introduction}
%\chapter{Introduction}
While it is generally understood that stars should accrete all of their mass during the Class 0 to Class II stages of protostellar evolution, the mass accretion timeline is not well-known. Observations of luminosities (and therefore mass accretion rates) in Class I and Class II young stellar objects (YSOs) are in tension with their expected main sequence masses \citep{Kenyon_LuminosityProblem_1990AJ, kenyon_hartmann_1995}. Typical mass accretion rates for low mass Class I/II YSOs are $10^{-8}-10^{-7} \ M_\odot$ yr$^{-1}$ \citep{Manara_PPVIIChapter_2023ASPC}. The lifetimes of disks around newly-formed YSOs are typically $< 10^7$ years, with decreasing accretion rates as the disk dissipates \citep{Michel_DiskLifetimes_2021ApJ}. Therefore, it is unlikely that steady accretion from the disk contributes much to the final mass of a young star.

The problem of when stars accrete their mass is at least partially solved by the existence of eruptive YSOs, which experience highly variable accretion in the form of outbursts of varying length and amplitude \citep{fischer_stellarMassAssembly_2023ASPC}. The most extreme of these are FU Ori objects (FUors), known to undergo large amplitude ($\Delta V = 4-5$ mag) optical and infrared photometric outbursts \citep{herbig_eruptive_1977}. The class is named for the prototype FU Orionis, which began its outburst in 1937 \citep{Wachmann_FUori_1954ZA} and has remained in its outbursting state for nearly 90 years. The long duration and high accretion rate \citep[$10^{-5}-10^{-4} \ M_\odot$ yr$^{-1}$, ][]{Nayakshin_TI_FUoriOutbursts_2024MNRAS} of FUor outbursts result in the disk transferring several $M_\mathrm{Jup}$ of material to the star in a single eruption. If all young stars experience FUor outbursts early in their formation, this would relieve the tension between observed accretion rates for Class I/II YSOs and their predicted main sequence masses. 

FUors have been suggested to in fact be similar to Class I/II YSOs based on their massive circumstellar envelopes and powerful outflows \citep{1994ApJ...424..793E, 2017MNRAS.468.3266R, 2023ApJ...945...80C, 2023A&A...672A.158S}. The detection of Herbig-Haro objects near FUors provides further evidence from their environments that their progenitors may be the younger Class I YSOs \citep{1985A&A...143..435R, StromStrom_V883OriDiscovery_1993ApJ, 1997AJ....114.2700R}. Recent work showing archival spectra of FUors pre-outburst has observationally confirmed that the progenitors to at least a few FUor outbursts are indeed Class I/II YSOs \citep{miller_evidence_2011,Hillenbrand_RNO54_letter_2023ApJ,2025A&A...695A.130S,Herczeg_Reipurth_HBC722_2025ApJ}. Spectral energy distribution (SED) fitting of the mid/far-infrared and millimeter photometry of several outbursting FUors has also shown their spectroscopic similarity to Class I/II YSOs at long wavelengths \citep{gramajo_spectral_2014}. As the extreme accretion luminosity ($L_\mathrm{acc}$) outshines the star by a factor of 100, at the shorter infrared and optical wavelengths only the hot inner disk is directly observable during outburst. It is thus challenging to conclusively characterize the underlying FUor stellar population. 
%Constraining whether any mass selection exists in which objects experience FUor outbursts is critical to understanding the ubiquity of the phenomenon.

Despite the impossibility of directly observing the central star, it is possible to infer the stellar mass from the spectrum of the disk. During outburst, the inner disk has an atmosphere that absorbs against the hot, viscously heated disk midplane \citep{Kenyon_FUori_disks_1988ApJ}, and superposes spectral absorption lines that can be studied at ultraviolet, visible and near-infrared (NIR) wavelengths. It is possible to model $M_*$ along with parameters such as $R_*$, $\dot{M}_{acc}$, $L_{acc}$, $T_{max}$,  by modeling the visible and NIR disk spectrum. However, doing so requires specific types of high-SNR, broad wavelength spectrophotometric data and difficult-to-obtain constraints on parameters like $A_V$, $d$, and $i$ \citep[e.g.,][]{Carvalho_V960MonPhotometry_2023ApJ, Carvalho_HBC722_2024ApJ}. 

The rotation curve of the outer disk is another plausible method for constraining the masses of FUors. One particularly bright and nearby FUor, V883 Ori, has an existing dynamical mass constraint from millimeter wavelength observations of gas emission in its outer disk \citep{Cieza_ALMA_V883Ori_2016Natur}. But thus far this is the only one.

Another strategy is to use only the common assumption of Keplerian rotation of the gas in the inner disk \citep{HartmannKenyon_V1057CygHighResolution_Disk_1987ApJ, welty_V1057_broadening_1990ApJ, welty_FUoriV1057CygDiskModelAndWinds_1992ApJ, Zhu_FUoriDifferentialRotation_2009ApJ}. 
Adopting the Keplerian rotation assumption, the rotation profile $v(r)$ is set by the mass of the central body. We thus can constrain the mass of a FUor by measuring $v(r)$, characterized by the velocity at the inner edge of the disk, $v_\mathrm{max}$. This fact has been used to break degeneracies in the SED modeling of FUors \citep[e.g.,][]{Carvalho_V960MonPhotometry_2023ApJ, Hillenbrand_RNO54_letter_2023ApJ}. However, $v_\mathrm{max}$ also depends on the inner truncation radius and the inclination of the disk, making directly mapping an observed $v_\mathrm{max}$ to the underlying stellar mass impossible. 

Rather than inferring a unique $M_*$ for each FUor based on a single rotational broadening measurement, our approach is to compare the distribution of rotational broadening measurements of FUors to that expected for different stellar mass distributions. This way we can probe the mass distribution of FUors at the population level. 
In this paper, we first endeavor to test whether the Keplerian assumption holds for the majority of the inner disk $r < 1$ au. Historically, line profile analysis of FUors has been limited to narrow wavelength ranges, spanning typically 100s of \AA\ in the visible and one or two echelle orders in the NIR. We expand our line profile analysis to the entire NIR from $1.0-2.4 \ \mu$m using high resolution spectra from the Keck/NIRSPEC instrument to clearly demonstrate the Keplerian rotation present in the disks of 20 FUors.

In Section \ref{sec:data}, we describe the Keck/NIRSPEC survey, our data reduction procedure, and additional archival spectra used in our mass distribution measurements. In Section \ref{sec:Disks}, we describe the implications of different parameters in a Keplerian disk across the near-infrared spectral range. In Section \ref{sec:CCFsAll}, we discuss the cross-correlation function (CCF) analysis technique we use to study the NIR line profiles in the sample, and compute CCFs to both show that the objects are in Keplerian rotation and measure the distribution of $v_\mathrm{max}$ values. In Section \ref{sec:simDist} we simulate the expected $v_\mathrm{max}$ distribution for different stellar mass distributions and compare them with the observed distribution. Finally, in Section \ref{sec:discussion}, we discuss the implications of our results for FUors generally and how future observations can better constrain the mass distribution of FUors.

\section{Sample and Data} \label{sec:data}

We obtained Keck/NIRSPEC \citep{McLean_nirspecDesign_1998SPIE} spectra for 23 FUors between 2022 and 2023. The sample was assembled after considering the ``Bona fide FUors'' and ``FUor-like'' sources in \citet{connelley_near-infrared_2018}, supplemented by new sources discovered in the years between that work and 2022: Gaia 17bpi \citep{hillenbrand_gaia_2018}, Gaia 18dvy \citep{2020ApJ...899..130S}, PGIR 20dci \citep{2021AJ....161..220H}, and RNO 54 \citep{Hillenbrand_RNO54_letter_2023ApJ}. We observed all sources north of declination $\delta>-35$ with brightness $K<14$, with a goal to obtain a per-pixel signal-to-noise ($SNR$) ratio greater than 30 in YJHK. Many of these objects are highly embedded, so the final wavelength coverage differs across the sample largely as a function of $A_V$. For 14 sources, we were able to obtain high resolution spectra across the entire desired YJHK range. For 5 sources we have data spanning JHK and for the remaining 2 we only have H and K band spectra. Thus for 90 \% of the sample we obtain high $SNR$ from at least $1.2-2.4 \ \mu$m and for 67 \% of the sample we have the entire $1.0-2.4 \mu$m. 

The spectra were all taken via standard ABBA nod sequences in generally good weather\footnote{The spectra taken in Dec 2022 were obtained during the peak of the eruption of Mauna Loa during that year. We were baffled by the highly variable sky background to the South until we realized we were seeing lava glow reflecting off the clouds. The temperature of the lava was exactly right to increase $K$ band sky backgrounds so that even the reddest objects were difficult to see in $K$ using the slit camera and we had to resort to finding the objects in $J$ band first.}. The spectra were extracted, deblazed, coadded, and telluric corrected using the \texttt{PypeIt} \citep{prochaska_pypeit_2020JOSS} module written for Keck/NIRSPEC and the telluric correction procedure described in \citet{Carvalho_PypeIt_2024RNAAS}. 

The telluric standards used were A0V stars that were easily accessible throughout the nights of the observing run. We observed standards at similar airmass to each of the sources (within $0.2$) to ensure accurate telluric removal. The hydrogen absorption lines in the standards were removed by first fitting a rotationally broadened $T_\mathrm{eff} = 10,000$ K, log$g=4.0$ BT-Settl \citep{2014IAUS..299..271A} model to the observed spectrum and dividing out the model. We omit spectral orders with significant telluric absorption, so for Y band we use orders $70-79$, for J band we use orders $57-66$, for H band we use orders $45-52$, and for K band we use orders $30-41$. 

We complement the NIRSPEC spectra 
with 15 $H$ and $K$ band spectra from the Immersion GRating INfrared Spectrometer  \citep[IGRINS,][]{Park_IGRINS_2014SPIE}, covering 7 sources from our NIRSPEC sample and adding $H$ and $K$ band spectra for 8 new sources. These 8 sources are FUors whose discovery was published following the completion of our NIRSPEC survey or which are too far to the south to have been observed as part of the survey: ESO H$\alpha$-148 \citep[Gaia 21elv;][]{2019MNRAS.486.4590C,2023MNRAS.524.3344N}, RNO 91 \citep[IRAS 16316-1540,][]{2021ApJ...919..116Y}, SPICY15470, SPICY35235, SPICY57130, SPICY68600, SPICY68696 \citep{2025ApJ...987...23C}, and WRAY 15-488 \citep{2019MNRAS.486.4590C}. The spectra were taken over the past 10 years at Gemini South, the McDonald Observatory Harlan J. Smith Telescope and the Lowell Discovery Telescope. The archived spectra were reduced by and are available at the Raw and Reduced IGRINS Spectral Archive \citep[RRISA,][]{RRISA_2025PASP}. As with our NIRSPEC spectra, we omit the bluest and reddest order of each band due to significant telluric absorption, so for both H and K band we use 19 spectral orders.

Our total sample for our Keplerian rotation analysis has coverage from $1.2-2.4 \ \mu$m for 19 FUors with NIRSPEC and $1.5-2.4 \ \mu$m for 8 FUors with IGRINS, while the sample for our $v_\mathrm{max}-$based mass distribution analysis has 28 FUors\footnote{We omit V1057 Cyg from the rotational broadening analysis because its absorption lines appear strongly wind-dominated and show little disk absorption, consistent with the findings of \cite{Szabo_V1057cyg_2021ApJ}}. The typical resolution $R \equiv \lambda/\Delta\lambda$ of the NIRSPEC spectra ranges from $R = 15000$ to $R = 25000$, depending on the choice of slit width, corresponding to a velocity resolution of $12-20$ km s$^{-1}$. The IGRINS spectra have a finer typical resolution of $R = 45000$, or a velocity resolution of $6$ km s$^{-1}$.

\section{The Spectral Implications of a Disk} \label{sec:Disks}

The idea that FUors have a thin, viscous accretion disk in approximately Keplerian rotation around the central star with $v\propto r^{-1/2}$ \citep{Kenyon_FUori_disks_1988ApJ} naturally implies that further away from the star, material will be orbiting more slowly. The temperature profile for the disk, $T\propto r^{-3/4}$, further implies that this slower rotating material should be cooler and hence radiate at redder wavelengths. 

\begin{figure}[!htb]
    \centering
    \includegraphics[width=0.99\linewidth]{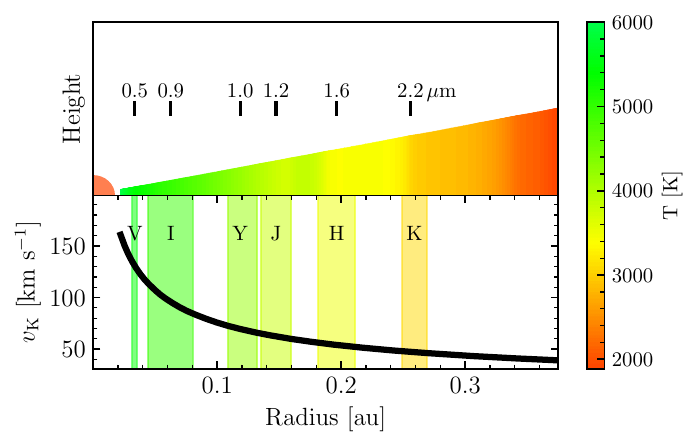}
    \caption{An illustration of the radial temperature and velocity structure of the disk of FU Ori. The vertical hashes in the upper panel mark the flux-weighted mean radii probed by the continuum at 0.5, 0.9, 1.0, 1.2, and 1.6 microns. The colors of the hashes correspond to the local temperature. 
    The colored bands in the lower panel mark the regions of the disk probed by the $V$, $I$, $Y$, $J$, and $H$, and $K$ photometric bands. Notice that the $H$ and $K$ band fluxes arise from radii greater than $0.2$ au ($40 \ R_\odot$). $H$ and $K$ band flux in disks with smaller $R_\mathrm{outer}$ values is dominated by the Rayleigh-Jeans tail of closer-in annuli, rather than spectra which peak at temperatures corresponding to $\sim 0.2-0.5$ au.} 
    \label{fig:cartoon}
\end{figure}

An illustration of the temperature structure of a FUor disk and the radial location probed by different wavelengths is shown in Figure \ref{fig:cartoon}. When looking at high resolution spectra of FUors, absorption lines at redder wavelengths should have narrower profiles than absorption lines at bluer wavelengths. The effect should also be seen in the excitation potential of the lines: features with greater excitation potential should probe hotter material and have broader absorption lines than features with lower excitation potentials. 

When modeling the high resolution visible range spectra of FU Ori, V1057 Cyg (pre-1994 fade event), and Z CMa, \citep{welty_FUoriV1057CygDiskModelAndWinds_1992ApJ} found the absorption lines in observed spectra did not follow this trend. They instead showed a lack of correlation between rotational line broadening and either wavelength or excitation potential. Disk model spectra successfully reproduced this lack of correlation, suggesting that the locations where lines form in the disk may not follow exactly the continuum emission location. 

However, \citet{herbig_high-resolution_2003} and \citet{PetrovHerbig_FUOriLines_2008AJ} later produced a disk model following a similar procedure to \citet{welty_FUoriV1057CygDiskModelAndWinds_1992ApJ}, and found their model did predict that line widths should correlate with both excitation potential and wavelength. \citet{Zhu_FUoriDifferentialRotation_2009ApJ} explained this discrepancy between disk model predictions by arguing that the temperature gradient in the region of the disk probed by visible range spectra depends on the choice of the maximum temperature in the disk, or $T_\mathrm{max}$. \citet{Zhu_FUoriDifferentialRotation_2009ApJ} also demonstrated that the high resolution spectrum of FU Ori at 5 $\mu$m has much narrower absorption lines than the visible range spectrum, as expected. 

\begin{figure}[!htb]
    \centering
    \includegraphics[width=0.99\linewidth]{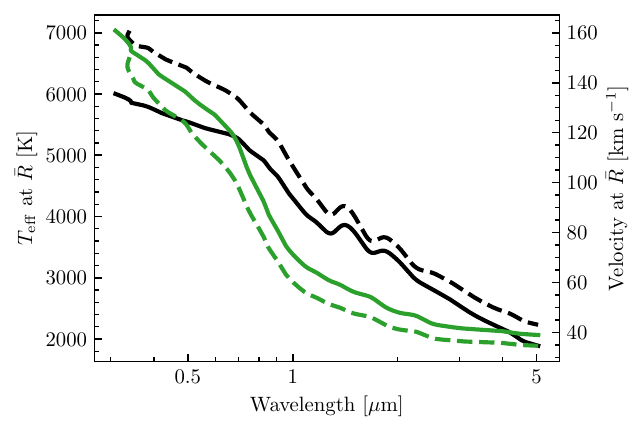}
    \includegraphics[width=0.99\linewidth]{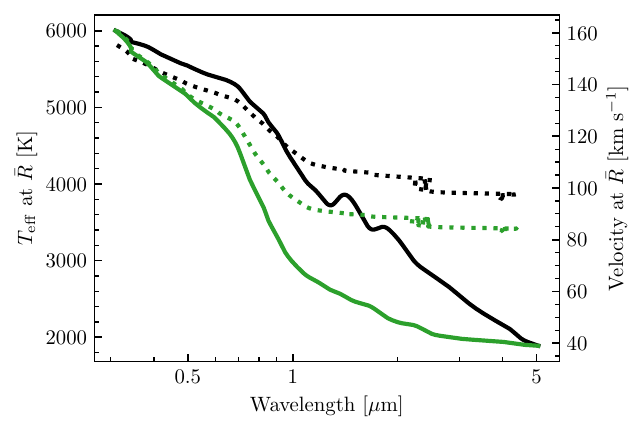}
    \caption{\textbf{Upper} The temperature (black lines) and velocity at the flux-weighted mean radius (green lines) as a function of wavelength for two disk models. The solid black and green lines shows the $T_\mathrm{max} = 6200$ K model, while the dashed lines show the $T_\mathrm{max} = 7800$ K model. \textbf{Lower:} Similar to the upper panel, but showing the impact of changing $R_\mathrm{outer}$ on the temperatures and velocities probed by each wavelength bin. The solid lines again show the fiducial model with  $R_\mathrm{outer} = 150 \ R_\odot$ (0.75 AU), while the dotted lines shows the $R_\mathrm{outer} = 25 \ R_\odot$ (0.15 AU) model. The oscillation in the temperature profile in the NIR is due to the fact that deep molecular absorption features probe cooler radii. }
    \label{fig:TempProf}
\end{figure}

\citet{carvalho_V960MonSpectra_2023ApJ} applied the same rotational broadening tests to V960 Mon and found that the widths of lines in visible range spectra do not correlate strongly with wavelength or excitation potential in that source, either. They then demonstrated that, as described in \citet{Zhu_FUoriDifferentialRotation_2009ApJ} for FU Ori, the $T_\mathrm{max}$ of the disk model impacts the dependence of visible range rotational broadening on the observed line width. Higher $T_\mathrm{max}$ values in the disk will cause the region from which the visible range spectrum emerges to have a greater radial $T_\mathrm{eff}$ gradient. Thus, as \citet{Zhu_FUoriDifferentialRotation_2009ApJ} concluded, searching for the predicted differential disk rotation in visible range spectra alone can yield inconclusive results. 

This effect is illustrated in the top panel of Figure \ref{fig:TempProf} for the disk model shown in Figure \ref{fig:cartoon} ($T_\mathrm{max} = 6200$ K) and a hotter model ($T_\mathrm{max} = 7800$ K). The temperature profiles show the flux-weighted temperature probed by each wavelength bin in the disk model, while the velocity profiles show the velocity at the flux-weighted mean radius ($\bar{R}$) of each bin. Notice the much steeper gradient in both velocity and temperature at visible wavelengths for the hotter model. The exact rotational broadening experienced by an individual line also depends on its excitation potential in the disk, not only the temperature probed by the continuum surrounding the line. 

Another important influence on the observed visible versus NIR line profiles of FUors is how the flux-weighted mean radii probed by different spectral regions of the disk model varies as a function of $R_\mathrm{outer}$, as shown in the bottom panel of Figure \ref{fig:TempProf}. For models with larger $R_\mathrm{outer}$, the typical radius probed by the 2 $\mu$m region of the spectrum is much further out in the disk than for smaller $R_\mathrm{outer}$. This is because the former has more low temperature contribution to the 2 $\mu$m from the disk atmosphere at large radii. The 2 $\mu$m contribution from a disk with very small $R_\mathrm{outer}$ will simply be that of the outermost annulus of the disk. Referring back to Figure \ref{fig:cartoon}, the spectrum of a disk that truncates at 25 $R_\odot$, or 0.15 AU, will show the same line broadening at $J$ and $K$ bands. Therefore, the differential rotation between 1 $\mu$m and 2 $\mu$m is much more extreme. The difference in velocity between $1-2 \ \mu$m for the $R_\mathrm{outer} = 25 \ R_\odot$ model is negligible compared to that in the $R_\mathrm{outer} = 150 \ R_\odot$ model. As is demonstrated in \citet{kospal_hbc722_2016A&A}, \citet{carvalho_V960MonSpectra_2023ApJ}, and \citet{Carvalho_HBC722_2024ApJ}, the $R_\mathrm{outer}$ can vary from object to object or even over time in one object. 

The above complications might suggest that measuring line widths at blue-optical or ultraviolet wavelengths should probe material close to $R_\mathrm{inner}$ and enable the most accurate measurement of the $v_\mathrm{max}$ in the disk from single line profiles. However, blueward of $\sim 5000$ \AA, the spectrum of FUors suffers heavy contamination from the strong outflow absorption near the star and the NUV/FUV spectrum is dominated by outflow emission \citep{Kravtsova_FUoriSTIS_2007AstL, Carvalho_FUVFUori_2024ApJ}. A further hindrance is the strong effect of extinction towards bluer wavelengths, making high signal-to-noise spectra of FUors difficult to obtain.

\section{The Power of Broad Spectrum Cross Correlation Analysis} \label{sec:CCFsAll}

We use the power of cross correlation techniques to measure differential rotation in near infrared spectra for the sample of 28 FUors. Despite some dependence of the measureable differential broadening on $R_\mathrm{outer}$, as discussed above, the effect is seen only over a large change in $R_\mathrm{outer}$. 

\subsection{Computing NIR CCFs of FUors} \label{sec:ComputeCCF}
We represent each NIRSPEC filter band ($Y$, $J$, $H$, and $K$) via a single cross-correlation function (CCF), so that four CCFs span the full $1.0 - 2.5$ $\mu$m spectral range of our spectra. 
When computing a CCF of FUor spectrum, the choice of reference spectrum is nontrivial. Due to blended contributions of different temperature components to a given wavelength range, the $T_\mathrm{eff}$ of a single-temperature reference spectrum will affect the weighting of absorption features and therefore the CCF profile. We avoid this by computing a reference accretion disk model spectrum based on FU Ori, which contains the appropriate mixed temperature contributions expected for an accretion disk. The details of the model can be found in \citet{Carvalho_HBC722_2024ApJ} and \citet{Carvalho_Thesis_2026}.

To begin, we adopt following physical parameters: $M_* = 1.0 \ M_\odot$, $R_\mathrm{inner} = 3.52 \ R_\odot$, $\dot{M} = 10^{-4.49} \ M_\odot$ yr$^{-1}$, $R_\mathrm{outer} = 150 \ R_\odot$. The resulting $T_\mathrm{max}$ of the model is 6200 K. Since we are continuum-normalizing the spectra, $A_V$ and distance are irrelevant here. The fiducial model that serves as our reference for the CCFs is computed with no rotational broadening ($i = 0^\circ$). In order to study the expected rotational broadening of the model absorption lines, we compute a grid of models with $i = 1^\circ - 89^\circ$. The resulting velocity range is $v_\mathrm{max} = 1 - 232$ km s$^{-1}$. We do not consider microturbulence in the model spectra, as we expect it to be $< 20$ km s$^{-1}$ and therefore below or comparable with the spectral resolution of most of our observations \citep{carvalho_V960MonSpectra_2023ApJ}. We discuss the impact of microturbulence on our measurements for the most face-on FUors in Section \ref{sec:discussion}. 

\begin{figure}[!htb]
    \centering
    \includegraphics[width=0.98\linewidth]{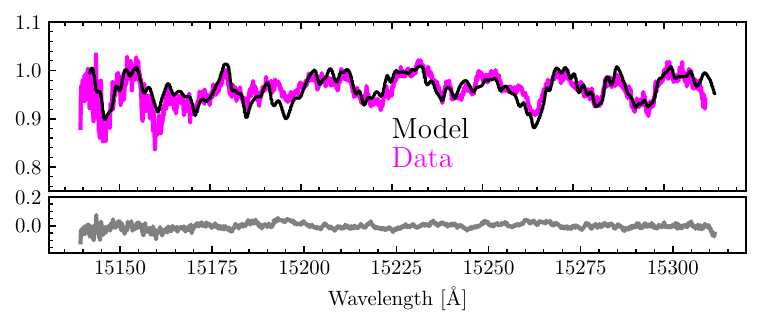}
    \includegraphics[width=0.98\linewidth]{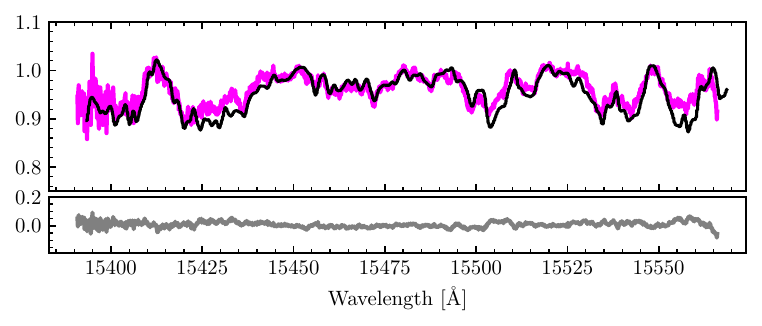}
    \includegraphics[width=0.98\linewidth]{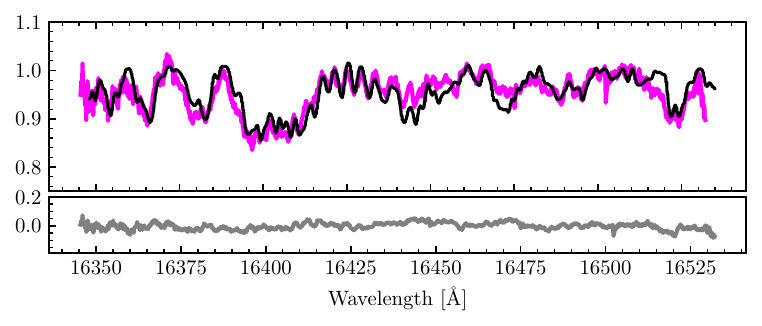}
    \includegraphics[width=0.98\linewidth]{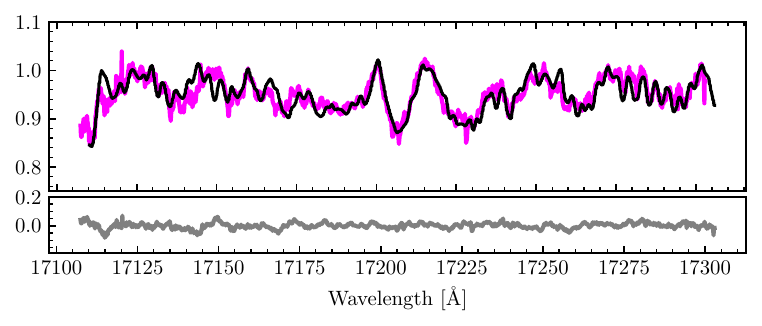}
    \caption{Several orders of the IGRINS spectrum of FU Ori (black) compared with the fiducial model, broadened to the best-fit $v_\mathrm{max} = 115$ km s$^{-1}$ for the source. Notice the excellent agreement with the data in both atomic and molecular features using only $v_\mathrm{max}$ as a free parameter in the model. The lower panel shows the residual between the model and the data.}
    \label{fig:IGRINSSpectra}    
\end{figure}

In Figure \ref{fig:IGRINSSpectra}, we present our fiducial model broadened to the best-fit $v_\mathrm{max}$ value of FU Ori derived in Section \ref{sec:HbandCCFs}.
Despite the higher $T_\mathrm{max}$ in the fiducial model versus the $T_\mathrm{max} = 6000$ K found for FU Ori \citep{Zhu_FUori_2007ApJ}, the model spectrum is almost identical to the IGRINS spectrum. We take this as validation of the relative insensitivity of the line profiles in the NIR to the $T_\mathrm{max}$ of a given FUor. We therefore proceed with our fiducial model grid with fixed $T_\mathrm{max} = 6200$ K for this analysis.

We then downsample the models to the wavelength sampling and resolution of the NIRSPEC or IGRINS spectrum in each of the available orders from $Y$ to $K$ band and compute the CCF between the reference model and each of the broadened models and the NIRSPEC spectrum. We combine the CCFs of each order to produce a geometric mean CCF following the guidance in \citet{Zucker_CombineCCFs_2003MNRAS}:
\begin{equation}
    CCF_\mathrm{mean} = \sqrt{ 1 - \left( \prod_1^{N}CCF_i^2 \right)^{1/N} } ,
\end{equation}
where $CCF_i$ is the CCF for a given spectral order and $N$ is the number of orders in a band used for the analysis. We omit the reddest and bluest order from each band due to heavy telluric correction. The resulting $CCF_\mathrm{mean}$ is a maximum-likelihood based combination of all $N$ orders, minimizing the effects of artifacts near the wings of any individual CCF. 

\subsection{The Wavelength Dependence of Rotational Broadening in FUors}

\begin{figure}[!htb]
    \centering
    \includegraphics[width=0.49\linewidth]{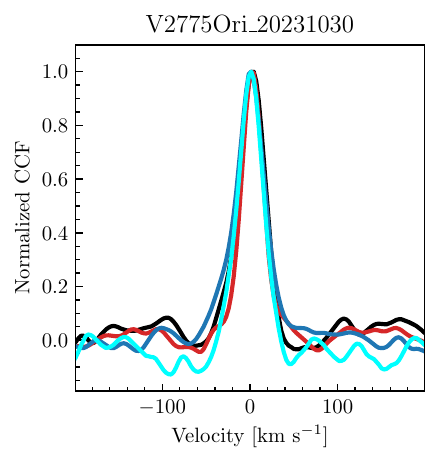}
    \includegraphics[width=0.49\linewidth]{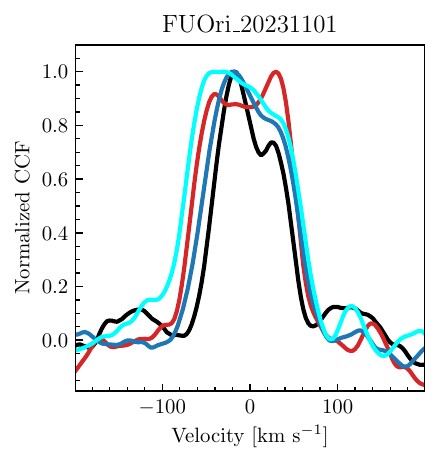}
    \caption{The CCFs of the Y (cyan), J (blue), H (red), and K (black) band NIRSPEC spectra for two objects: V2775 Ori (left panel) and FU Ori (right panel). V2775 Ori is a narrow-line FUor, with a $v_\mathrm{max}$ measurement limited by the instrument broadening of NIRSPEC. FU Ori represents the typical object in the FUor population, with a $v_\mathrm{max} \sim 115$ km s$^{-1}$ (see Section \ref{sec:HbandCCFs}). The rest of the sources are shown in the full figure set available in the online journal. }
    \label{fig:CCF_Multiband}
\end{figure}

The $CCF_\mathrm{mean}$ curves for two example sources are shown in Figure \ref{fig:CCF_Multiband}. In the majority of FUors in the sample, the $K$ band CCFs are clearly narrower than the $J$ or $Y$ band CCFs. There are some objects, however, in which the expected narrowing of the CCF at longer wavelengths is not seen. Comparison with the model CCFs, however, clarifies why even in the NIR the differential rotation may appear as strongly as expected. 

\begin{figure}[!htb]
    \centering
    \includegraphics[width=0.95\linewidth]{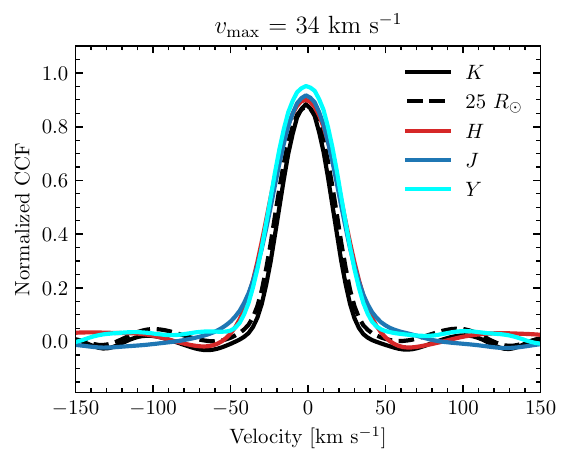}
    \includegraphics[width=0.95\linewidth]{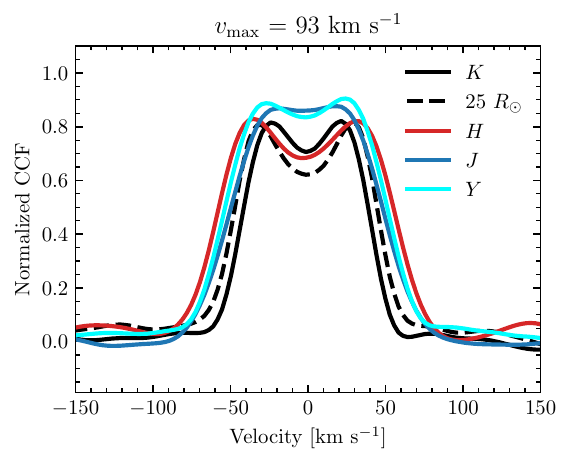}
    \includegraphics[width=0.95\linewidth]{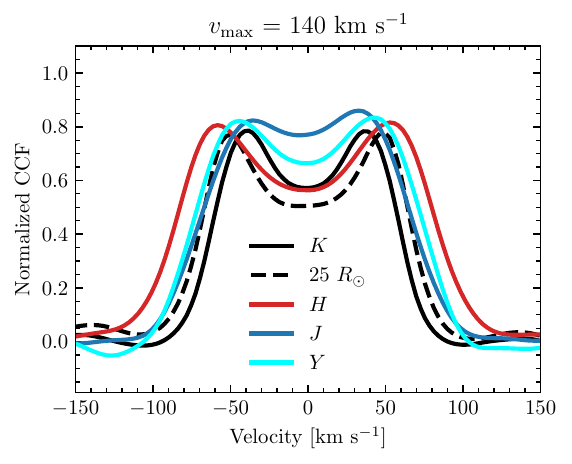}
    \caption{The CCFs produced from disk models with $R_\mathrm{outer} = 150 \ R_\odot$ of the Y (cyan), J (blue), H (red), and K (black) band NIRSPEC spectra. The brown dashed line shows the K band CCF produced from a disk model with $R_\mathrm{outer} = 25 \ R_\odot$. Notice that the expected narrowing of line widths at longer wavelengths is stronger for larger $R_\mathrm{outer}$ and becomes negligible for small $R_\mathrm{outer}$ values. Different panels show the results for different values of $v_{max}$}
    \label{fig:CCFModels}
\end{figure}

In Figure \ref{fig:CCFModels}, we show three sets of model CCFs (from the grid computed in Section \ref{sec:HbandCCFs}) for different $v_\mathrm{max}$ values and the four bands. In all three model sets, for the fiducial $R_\mathrm{outer} = 150 \ R_\odot$, the $K$ band CCFs are indeed predicted to be narrower, while the models with $R_\mathrm{outer} = 25 \ R_\odot$ have $K$ band CCFs as broad as the bluer bands. This can be explained by the sensitivity of the flux-weighted mean radius probed at each wavelength to the $R_\mathrm{outer}$ value of a given FUor, as discussed in Section \ref{sec:Disks}.

\subsection{H Band as the Ideal Wavelength Range Probing Rotational Broadening} \label{sec:HJustification}
The technique we present in Section \ref{sec:HbandCCFs} is focused on $H$ band, rather than $Y$, $J$, or $K$. While the model results presented in Section \ref{sec:DifferentialRot} demonstrate that this technique should work in any band, there are several reasons that, in practice, $H$ band is optimal for the measurement.

The primary benefit of $H$ band is that it is far enough in the infrared that even for deeply embedded sources we are able to obtain a high sensitivity and high resolution spectrum. This ensures that for the majority of known FUors, it will be possible to obtain a $v_\mathrm{max}$ measurement using existing facilities. Despite the fact that $K$ band is redder still, it is not ideal for this measurement due to its relative paucity of atomic features \citep{Rayner_IRTF_2009} and potential for strong infrared veiling from hot dust \citep{carvalho_V960MonSpectra_2023ApJ,Carvalho_HBC722_2024ApJ}. Furthermore, the strongest and most line-rich molecular features in $K$ band, the CO $(\Delta \nu = 2)$ ro-vibrational transitions, have been shown to be strongly impacted by outflow absorption \citep{Carvalho_HBC722_2024ApJ}. Thus, $K$ band is not a reliable wavelength range for this measurement.

For bluer bands, a similar problem to that of the $K$ band CO arises: outflow absorption. In both $Y$ and $J$ bands, there are several strong features that are sensitive to outflows, such as the \ion{He}{1} 1.083 $\mu$m line \citep{connelley_near-infrared_2018}, as well as less-appreciated features like \ion{Sr}{2} 1.033 and the several \ion{Si}{1} features across $J$ band. The NIRSPEC data we present in this paper will enable future characterization of the most wind-sensitive features, which can be used to inform the lines to be masked for a $v_\mathrm{max}$ measurement made from $Y/J$ bands. In the context of this study, however, we have $Y$ and $J$ band spectra at sufficient sensitivity to apply this analysis to only half of the sources in the sample.

\subsection{Using CCFs in H band to Measure $v_\mathrm{max}$} \label{sec:HbandCCFs}

An effective means of using the high resolution spectrum of a FUor to constrain SED fits has been to measure the Keplerian rotational broadening in the spectrum. Recall that if the material is in a Keplerian orbit around the central star, then its velocity at a given radius, $r$, is $v_\mathrm{Kep}(r) = \sqrt{GM_*/r}$. The velocity we observe is projected along our line of sight according to the inclination, $i$, of the disk, so we observe $v_\mathrm{Kep}(r)\sin i$. The maximum velocity we observe is at $r = R_\mathrm{inner}$, and is given by $v_\mathrm{max} = \sqrt{GM_*/R_\mathrm{inner}} \sin i$. The fact that $v_\mathrm{max}$ contains $M_*$, $R_\mathrm{inner}$ and $i$ in its expression gives it the power to constrain these parameters in the SED model. 

To obtain estimates of $v_\mathrm{max}$ that are minimally impacted by absorption/emission from outflows and to account for the Keplerian velocity profile in the disk, we propose a new, CCF-based technique using high resolution $H$ band spectra. As mentioned in Section \ref{sec:HJustification}, the $H$ band is ideal due to its mix of atomic and molecular spectral features, and line absorption profiles that are relatively insensitive to the maximum temperature in the disk. 

The technique relies on comparing the grid of model H band CCFs, which span a wide range of $v_\mathrm{max}$ values, with the $CCF_\mathrm{mean}$ obtained for a given object. Our metric of comparison is the reduced $\chi^2$, which we minimize on the grid of CCFs to identify the correct best-fitting $v_\mathrm{max}$ value for a system. The $H$ band $CCF_\mathrm{mean}$ for each source in the sample is shown along with its best-fit model $CCF_\mathrm{mean}$ in Appendix \ref{app:AllMeasurements}.

Using this technique, we are able to measure the $v_\mathrm{max}$ values for all of the objects in our Keck/NIRSPEC survey and those with archival IRTF/IGRINS spectra with coverage in $H$ band. 
We note that the resolution of the NIRSPEC spectra is $\sim 15,000$, so the instrumental line broadening is 20 km s$^{-1}$. The IGRINS spectra have $R\sim 45,000$, with a corresponding line broadening of 6.67 km s$^{-1}$. This places a systematic upper bound on the measurable $v_\mathrm{max}$ in certain objects including V1515 Cyg, V2775 Ori, and PP 13S. 

\begin{figure*}
    \centering
    \includegraphics[width=0.9\linewidth]{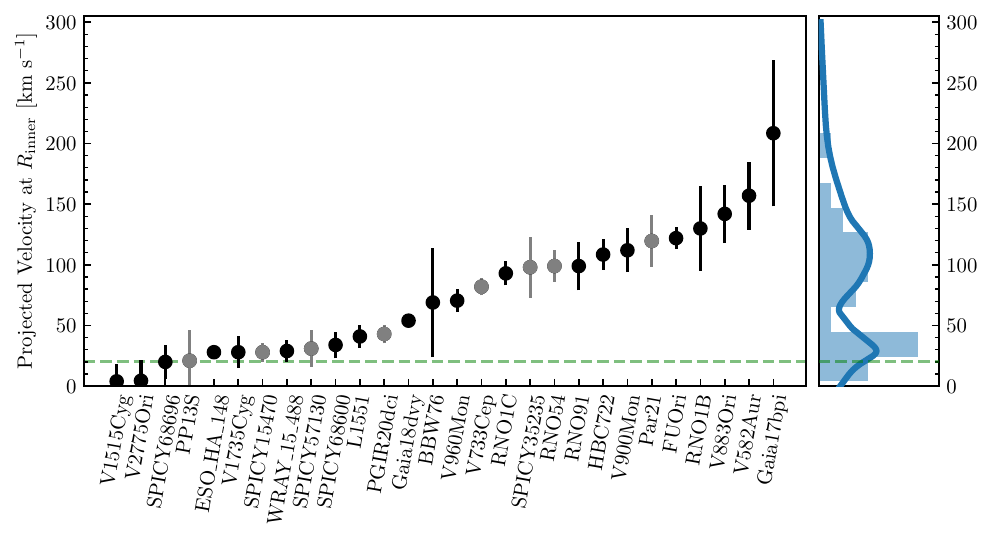}
    \caption{The measured $v_\mathrm{max}$ values for each object in the sample. \textbf{Left:} The $v_\mathrm{max}$ values for each object, shown with error bars. 
    Measurements marked in grey are optically faint, with $g > 20$ mag, and therefore difficult to analyze at shorter wavelengths. The green horizontal line marks the NIRSPEC resolution limit. \textbf{Right:} The binned histogram and Gaussian smoothed histogram of the $v_\mathrm{max}$ measurements.}
    \label{fig:VmaxVals}
\end{figure*}

The $v_\mathrm{max}$ measurements and their uncertainties are summarized in Figure \ref{fig:VmaxVals}. For objects observed over multiple epochs, we averaged their $v_\mathrm{max}$ measurements and combined their uncertainties in quadrature. 
The individual measurements for each spectrum are given in Appendix \ref{app:AllMeasurements}, as well as figures showing both data and best-fit CCFs alongside the $\chi^2$ values as a function of $v_\mathrm{max}$.

\section{Simulating the $v_\mathrm{max}$ distribution of FUors} \label{sec:simDist}

As can be seen in Figure \ref{fig:VmaxVals}, the distribution of $v_\mathrm{max}$ values in the sample has two distinct peaks: one at $\sim 30$ km s$^{-1}$ and another at $\sim 120$ km s$^{-1}$. To first order, one might imagine that the bimodality could arise from bias in the inclination distribution. However, two points negate this. The first is that HBC 722 and V960 Mon have similar $v_\mathrm{max}$ values despite HBC 722 being relatively edge-on \citep[$i > 75^\circ$][]{kospal_hbc722_2016A&A,Carvalho_HBC722_2024ApJ} and V960 Mon being face-on \citep[$i < 20^\circ$][]{Kospal_ALMA_2021, carvalho_V960MonSpectra_2023ApJ}. The second is less direct and requires assuming that a source being optically faint can be treated as a proxy for it being more edge-on than face-on. Filtering sources that have $g > 20$ mag, which are difficult to detect with existing visible range surveys, shows they are uniformly distributed across the $v_\mathrm{max}$ distribution. 

There are three important physical quantities that impact an individual observed $v_\mathrm{max}$: $M_*$, $R_\mathrm{inner}$, and $\sin i$, each of which has its own underlying distribution in the sample. To understand the origin of the bimodality in the $v_\mathrm{max}$ distribution, we must simulate the $v_\mathrm{max}$ measurements under different assumed mass functions and mass-radius relations. 

%Here we describe the procedure we follow 
To simulate the $v_\mathrm{max}$ distribution, we first assume an underlying mass function for the FUors, from which we draw 10,000 random samples. Then, we assign a radius to each stellar sample using an adopted isochrone, and follow the usual assumption that during outburst, $R_\mathrm{inner} \sim R_*$ \citep{Kenyon_FUori_disks_1988ApJ}. We then assign a randomly drawn inclination (uniformly sampled from $i = 1^\circ$ to $i = 89^\circ$) to each star and compute $v_\mathrm{max,sim} = \sqrt{G M_{*,\mathrm{sim}}/R_\mathrm{inner,sim}} \sin i_\mathrm{sim}$ for each sample sim. 

We broaden our fiducial high-resolution spectrum model to each $v_\mathrm{max,sim}$ by interpolating over the model grid used in Section \ref{sec:HbandCCFs}. We then add Gaussian noise with a standard deviation equal to $10\%$ of the continuum level to each simulated spectrum, apply an assumed 15 km s$^{-1}$ instrumental broadening, and compute the CCF. We use the procedure in Section \ref{sec:HbandCCFs} to measure the best-fit $v_\mathrm{max}$ for each sample and assign an uncertainty to the measurement. 
Using these synthesized measurements, we construct both a Gaussian Smoothed Histogram \citep[GSH,][]{Roychowdhury_FUoriV883OriDist_2024RNAAS} and a cumulative distribution function (CDF) to compare with our H band measurements for the FUor population.

We test two different possible initial mass functions for FUors: the solar-neighborhood IMF reported by \citet{Kirkpatrick_IMF_2024ApJS}, K24 hereafter, and a uniformly distributed (``flat'') IMF. We chose the K24 IMF for its completeness in the brown dwarf (BD) mass range. To test whether the $v_\mathrm{max}$ distribution is sensitive to BD FUors, we test the case where we use the entire K24 IMF (for $M_* < 2 \ M\odot$) and one where we restrict the K24 IMF to the lower range of stellar masses, so we sample only $0.1-2 \ M_\odot$. In the latter case, the K24 IMF is consistent with others that have power law indices $\sim 2.3$ \citep[e.g.,][]{Kroupa_IMF_OG_2001, Chabrier_IMF_2003PASP}. We restrict the flat IMF to the same $0.1-2 \ M_\odot$ range for consistency. 

The isochrone we adopt is the 1 Myr PARSEC v1.2 isochrone \citep{PARSEC_v1p2_2014MNRAS}. For the IMF that includes BDs, we supplement the PARSEC radii with radii from the \citet{Baraffe_isochrones_2015A&A} 1 Myr isochrone, which reaches $M_* = 0.01 \ M_\odot$. We also ran simulations using the Feiden et al. (2016) magnetic isochrones and did not find a significant difference in the predicted $v_\mathrm{max}$ distribution.

\begin{figure}[!htb]
    \centering
    \includegraphics[width=0.99\linewidth]{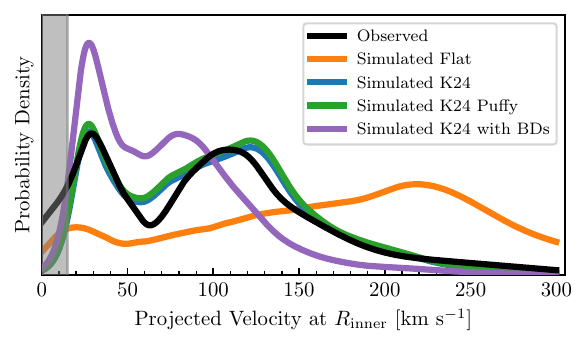}
    \includegraphics[width=0.99\linewidth]{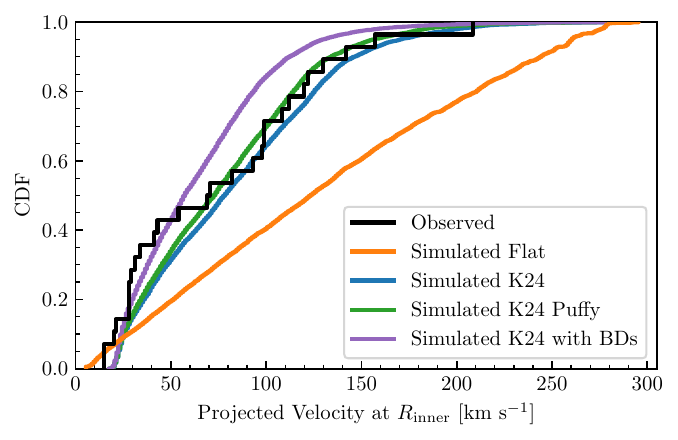}
    \caption{The GSHs (upper panel) and CDFs (lower panel) of the measured FUor $v_\mathrm{max}$ values and our 4 simulations. The GSHs and CDFs of the K24 and K24 Puffy simulations are near-perfect morphological matches to the observations, while the K24 with BDs and Flat simulations overpredict low $v_\mathrm{max}$ and high $v_\mathrm{max}$ values, respectively. }
    \label{fig:CCF_dists}
\end{figure}

The GSHs and CDFs of the simulated $v_\mathrm{max}$ distributions are shown alongside the observed GSH and CDF in Figure \ref{fig:CCF_dists}. Qualitatively, it is immediately clear that the simulated K24 IMF that includes BDs in the sampled mass range overproduces sources at very low $v_\mathrm{max}$ relative to the observed sample. The simulated flat IMF, on the other hand, overproduces sources at very high $v_\mathrm{max}$ relative to the observed sample. The K24 IMF that is limited to $0.1-2 \ M_\odot$ matches the observations well. 
%The implications of these observations are discussed in Section \ref{sec:IMFdisc}. 
Notice also that all of the simulations reproduce the overdensity of sources at $v_\mathrm{max} = 20-30$ km s$^{-1}$, which is the result of the line broadening profiles of NIRSPEC and IGRINS.

We conclude that a flat mass distribution is inconsistent with the known sample of FUors. We further find that an IMF like that describing the current solar neighborhood provides a reasonable match to the double-peaked distribution of measured $v_{max}$ values for the FUors. The match is less good when brown dwarfs are included in the solar neighborhood comparision,
likely due to incompleteness of the known FUor sample into the brown dwarf regime. 

We added a simulation case where we assume that the FUors are uniformly ``puffy'' relative to typical YSOs by increasing the radii of the sampled objects by 20\%. This is intended to test the long-held expectation that the high accretion rate of FUors should heat the star itself sufficiently to increase their radii during and following the outburst \citep{hartmann_fu_1996}. The expectation is supported by the large estimated stellar radius of FU Ori, for which $R_* \sim R_\mathrm{inner} =3.5 \ R_\odot$ \citep{Perez_FUoriALMA_2020ApJ}. The mass of FU Ori is estimated to be $M_* = 0.6 \ M_\odot$ \citep{Perez_FUoriALMA_2020ApJ}, so its radius is more than 50 \% greater than that predicted for a 0.5 Myr system\footnote{Although the age of FU Ori is not known, its SED is consistent with that of a Class II YSO, suggesting it is unlikely to be much younger than 1 Myr.} \citep{PARSEC_v1p2_2014MNRAS}. However, not all FUors appear to have such significantly inflated radii. For example, the estimated radius of V960 Mon is approximately consistent with the predictions of 1 Myr isochrones \citep{Carvalho_V960MonPhotometry_2023ApJ}. Unfortunately, in the simulated $v_\mathrm{max}$ distributions there is little difference between the use of PARSEC-derived radii and the ``puffy'' radii. 

\section{Summary and Discussion} \label{sec:discussion}

We have used broadband YJHK spectra at high resolution of FUors first to demonstrate the differential rotation in their disks,
and second to use $v_{max}$ measurements to probe the underlying mass distribution of their central stars. Below, we discuss the consequences of these measurements in the context of existing FUor literature and future studies. 

\subsection{Differential Rotation in FUor Disks} \label{sec:DifferentialRot}

The broad wavelength coverage in our NIRSPEC data enable us to clearly detect differential rotation in the disks of several FUors by showing that the Keplerian line broadening is weaker in $K$ band (2.2 microns) than in $Y$ band (1 micron). We are able to reproduce the observed decreasing broadening as a function of wavelength with a thin, viscous accretion disk in Keplerian rotation. However, there are sources whose spectra do not reflect the expected differential broadening between bluer and redder bands. Upon investigation of how different disk model parameters can impact the observed rotational broadening, we found that varying $R_\mathrm{outer}$ parameter enabled us to reproduce the consistent line widths between the bluest and reddest NIR bands. 

These results, combined with the work of \citet{carvalho_V960MonSpectra_2023ApJ} in visible range spectra, highlight the peril of mapping individually measured line widths to line formation locations in the disk. Although the disks do show clear Keplerian rotation profiles, diagnosing the profile from individual lines suffers confusion from varying dependence on $T_\mathrm{max}$ and the fact that for small $R_\mathrm{outer}$ values, low velocity annuli of the disk are not probed by the NIR spectrum. For most of the sources in our survey, the line narrowing between $Y$/$J$ band and $K$ band is unambiguous and clearly matches that predicted in our fiducial model. For others, a broader wavelength comparison will be needed to reveal the differential rotation, likely spanning $0.5 \ \mu$m to $3.5 \ \mu$m, as shown by \citet{Zhu_FUoriDifferentialRotation_2009ApJ}.

The differing reliability of a given wavelength range to detect the differential rotation in the disk is a feature of the disk model and arises naturally when considering disks with a variety of physical parameters. In our Section \ref{sec:DifferentialRot} analysis, V960 Mon shows no differential rotation measured between $Y$ and $K$ band, though it is a well-known bonafide FUor. However, the source is known to have a small $R_\mathrm{outer} \sim 25 \ R_\odot$ \citep{Carvalho_V960MonPhotometry_2023ApJ}, which we demonstrated in Section \ref{sec:DifferentialRot} can flatten the line broadening versus wavelength relation. It is thus crucial to interpret line profiles of FUors in tandem with a range of disk models, accounting for the range of physical properties observed in the sample. While knowing what $R_\mathrm{outer}$ to use in the models a priori is challenging, there are means of estimating it based on NIR colors, as discussed in \citet{2024ApJ...976L...5C}, since disks with small $R_\mathrm{outer}$ tend to be bluer than those with large $R_\mathrm{outer}$.

For the 8 FUor candidates observed exclusively by IGRINS, we sought to detect differential rotation between their H and K band spectra. However, ESO-H$\alpha$-148, SPICY15470, SPICY57130, SPICY68600, SPICY68986, and WRAY 15-488 are all narrow-lined and below the resolution limit of IGRINS to detect their differential rotation. RNO 91 and SPICY 35235 should have broad enough spectra lines to detect their differential rotation, but do not show it. As discussed above, however, this is not necessarily disqualifying. Further analysis of these sources is necessary, but given their FUor-like lightcurves and spectroscopic similarity to other narrow-lined FUors like V1515 Cyg, we adopt these into the sample as confirmed FUors. 

\subsection{The Mass Distribution of FUors} \label{sec:IMFdisc}

Prior to this work, only 11 FUors possessed published mass estimates: 
FU Ori \citep[0.6 $M_\odot$][]{Perez_FUoriALMA_2020ApJ}, V883 Ori \citep[1.3 $M_\odot$][]{cieza_v883Ori_2018MNRAS}, V960 Mon \citep[0.6 $M_\odot$][]{Carvalho_V960MonPhotometry_2023ApJ}, RNO 54 \citep[0.23 $M_\odot$][]{Hillenbrand_RNO54_letter_2023ApJ}, HBC 722 \citep[0.2 $M_\odot$][]{Carvalho_HBC722_2024ApJ}, Gaia 20bdk \citep[2.7 $M_\odot$][]{2025A&A...695A.130S}, V890 Aur \citep[0.17 $M_\odot$][]{Hillenbrand_V890Aur_2025}, BBW 76 (0.2 $M_\odot$), V1057 Cyg (1.5 $M_\odot$), V1515 Cyg (0.8 $M_\odot$), and Gaia 17bpi (0.1 $M_\odot$) \citep{Carvalho_Thesis_2026}. 
The technique presented in this paper provides a means of probing the mass distribution of FUors based on a sample of 28 objects. With this larger sample, we are able to demonstrate that the FUor mass distribution appears to match the stellar mass distribution in the solar neighborhood well. 

Our distribution of $v_\mathrm{max}$ measurements for FUors is the first of its kind and reveals the importance of high resolution NIR spectroscopy of these sources. \citet{Carvalho_Thesis_2026} demonstrates the criticality of the $v_\mathrm{max}$ value for constraining SED fits to FUors to precisely measure physical parameters of individual sources. We further showed here that the $v_\mathrm{max}$ distribution for an ensemble of sources can be used to probe the underlying mass distribution of FUOrs, and potentially probe the mass-radius relation of FUors. 

In Section \ref{sec:simDist}, we showed that the different assumed mass distributions: the K24 IMF with BDs, without BDs, and the flat IMF produce different structures in the simulated $v_\mathrm{max}$ distribution. Despite the limited sample size, we are able to distinguish a flat IMF from a more physical IMF like that of K24, and find that the FUors appear to arise from an IMF similar to that in the solar neighborhood.

Critically, there is no bias toward high mass stars undergoing FUor outbursts, as demonstrated by the poor agreement with even a uniform mass distribution. 
Another important, though marginal, trend is the apparent incompatibility with the presence of BDs in the FUor sample. However, it is possible this is due to observational bias. The low masses of BD FUors would likely correlate with lower mass accretion rates, combining to produce very low luminosities. This is the case in the two lowest mass FUors discovered to date, Gaia 17bpi \citep[$L_\mathrm{acc} \sim 6 \ L_\odot$,][]{Carvalho_Thesis_2026} and V890 Aur \citep[$L_\mathrm{acc} \sim 8.4 \ L_\odot$][]{Hillenbrand_V890Aur_2025}. Such low outburst luminosities and the expected blue color of the outbursting disk \citep[if $T_\mathrm{max} \sim 5000-7000$ K, as is the case for most FUors,][]{hartmann_fu_1996}, makes these objects difficult to distinguish from more massive non-outbursting YSOs photometrically. New high sensitivity infrared photometric and spectrophotometric surveys are opening up a new frontier in FUor detection, however, and may soon lead to the first discovery of a BD FUor \citep{2026ApJ...997..263F, 2026ApJ..1000L...8K}.

%All-sky spectroscopic surveys like SPHEREx \citep{2025arXiv251102985B} may enable detecting candidate BD FUors for more detailed follow-up and confirmation.  \lahcomm{do we need this gratuitous sentence? what is the argument here??  i think you should model a 0.03 Msun source and see what the predicted spectrum looks like at spherex resolution. i am guessing it isn't so distinct. }

Finally, we note that our $v_\mathrm{max}$ measurement technique requires only a high resolution $H$ band spectrum of an FUor with moderate signal-to-noise. Obtaining these for the remaining $\sim 40$ FUors in the known sample \citep{CCP_OYCAT_2025JKAS} is possible with the instruments currently available, but requires large apertures for the fainter sources, particularly those discovered at infrared wavelengths in the Southern sky \citep[e.g.,][]{Guo_VVV_FUoriBursts_2024MNRAS}. Adding $v_\mathrm{max}$ measurements for those sources will improve our ability to further constrain the details of the FUor mass distribution (see Appendix \ref{app:SampSize}). It will also enable better-constrained SED fitting, which is crucial for precise measurements of the physical parameters of individual sources in the FUor sample.

\subsection{Interpreting Lower Limits On FUor Line Widths}
The pile-up of sources seen at low $v_\mathrm{max}$ is well-reproduced by our simulations and suggests that it arises from a bias in the measurement method due to the line-broadening functions of NIRSPEC and IGRINS. As can be seen in Figure \ref{fig:CCF_dists}, this impacts the ability to measure the $v_\mathrm{max}$ values of both stellar-mass face-on FUors and BD-mass inclined FUors. Higher resolution spectrographs are necessary to measure the Keplerian broadening in these sources.

Besides the lower limit on measureable $v_\mathrm{max}$ set by instrument resolution, there is an astrophysical limit as well. Microturbulence is predicted in FUor disks at the several km s$^{-1}$ level \citep{Zhu_outburst_FUOri_2020MNRAS}, particularly given the extremely high viscosity driven by the magneto-rotational instability. The high resolution visible range spectra of FUors, including that of FU Ori itself, have absorption line profiles that are flatter and more box-like near the line centers than is expected from the Keplerian disk model \citep{herbig_high-resolution_2003,PetrovHerbig_FUOriLines_2008AJ}. \citet{carvalho_V960MonSpectra_2023ApJ} found that the inclusion of $\sim 20$ km s$^{-1}$ of non-Keplerian line broadening successfully reproduced the flat line profiles, which can be attributed to the broadening to radial and meridional motion of turbulent eddies in the disk. For face-on or very low mass FUors, microturbulence may dominate the broadening and make measuring the Keplerian $v_\mathrm{max}$ impossible.

Future observations of narrow line FUors with extremely high resolution spectrographs like CRIRES on the Very Large Telescope and HISPEC on Keck II will be crucial to disentangling the systematic and astrophysical lower bounds to the $v_\mathrm{max}$ distribution. They will also enable us to better probe, at a population level, the potential existence of BD FUors.

\section{acknowledgments}
ASC thanks the staff at Keck Observatory for their support during the data collection for this survey, and in particular Greg Doppmann, for his expertise and guidance on optimal observing with the NIRSPEC instrument. 
ASC also thanks Joshua Lovell, Sean Andrews, David Wilner, and Jessie Miller for many helpful discussions during the course of this work. 

This work used The Immersion Grating Infrared Spectrometer (IGRINS) developed under a collaboration between the University of Texas at Austin and the Korea Astronomy and Space Science Institute (KASI) with the financial support of the US National Science Foundation under grants AST-1229522, AST-1702267 and AST-1908892, McDonald Observatory of the University of Texas at Austin, the Korean GMT Project of KASI, the Mt. Cuba Astronomical Foundation and Gemini Observatory.
The RRISA is maintained by the IGRINS Team with support from McDonald Observatory of the University of Texas at Austin and the US National Science Foundation under grant AST-1908892.
%\end{acknowledgments}

\software{\texttt{Astropy} \citep{astropy_2013,astropy_2018,astropy_2022}, \texttt{NumPy} \citep{harris2020array}}

%\section{Acknowledgements}
%\begin{acknowledgements}
%The authors thank Antonio Rodriguez for insightful conversations and suggestions. 
%\end{acknowledgements}

\bibliography{references}{}
\bibliographystyle{aasjournal}

\appendix 

\restartappendixnumbering

\section{All $v_\mathrm{max}$ Measurements and Accompanying Plots} \label{app:AllMeasurements}

The individual $v_\mathrm{max}$ measurements for each NIRSPEC and IGRINS spectrum in the sample are reported in Table \ref{tab:VelsRot}. The order of sources in the table is arbitrary, though when we have multiple observations of each source, they are sorted by date.
%\lahcomm{what is the order of this table??  do we need one?}

\begin{figure}[!htb] \label{fig:CCF}
    \centering
    \includegraphics[width=0.98\linewidth]{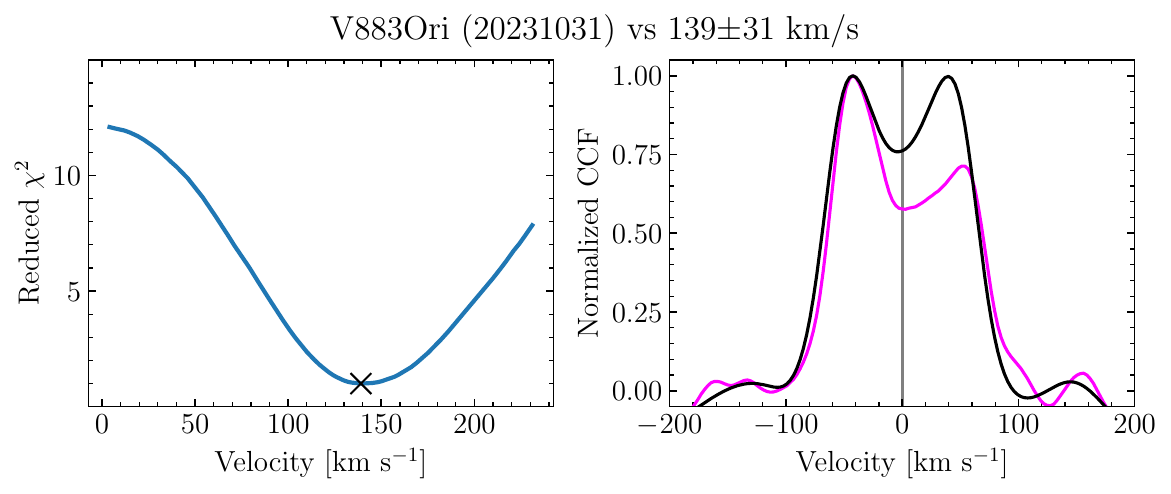}
    \caption{The results of the $H$ band CCF best-fit used to measure the rotational broadening in V883 Ori. 
    \textbf{Left:} The reduced $\chi^2$ values for the grid of model CCFs and the V883 Ori CCF. The best-fit value of $v_{max}$, as determined from the minimum $\chi^2$, is marked with a $\times$ symbol. \textbf{Right:} The mean CCF from 10 H band orders of the NIRSPEC spectrum of V883 Ori (magenta) and the best-fitting mean CCF from the model grid, corresponding to a $v_\mathrm{max} = 139$ km s$^{-1}$. The stronger blue-shifted absorption in V883 Ori is likely due to a rotating disk wind traced by molecular bands like CO (3-0).  }
    \label{fig:CCFs_V883Ori}
\end{figure}

As an example of our best-fit broadening procedure, the $\chi^2_r$ curve for the source V883 Ori, alongside the $CCF_\mathrm{mean}$ for its H band spectrum is shown in Figure \ref{fig:CCFs_V883Ori}, compared with the model $CCF_\mathrm{mean}$ for the best-fit value of $v_\mathrm{max} = 139$ km s$^{-1}$. Figure \ref{fig:CCFs_V883Ori} is presented as a figure set showing analogous plots for each $v_\mathrm{max}$ measurement shown in Table \ref{tab:VelsRot} and can be viewed in its entirety on the online journal. 

Some of the target CCFs show an asymmetry between strength of the red peak of the CCF and the blue peak. This has been observed in visible range analysis of disk-tracing line in FU Ori and was attributed to small azimuthally-confined surface density inhomogenieties in the disk orbiting near the star \citep{2012MNRAS.426.3315P}. When the blue side is particularly strong relative to the red, there may be significant wind absorption in the CO $\Delta \nu = 3$ bands, as was demonstrated in the source HBC 722 \citep{Carvalho_HBC722_2024ApJ}.

\startlongtable
\begin{deluxetable}{ccccc}
\tablecaption{The $v_\mathrm{max}$ measurements for confirmed and high-likelihood candidate FUors with a Keck/NIRSPEC $H$ band spectrum. 
\label{tab:VelsRot}}
\tablewidth{0pt}
\tablehead{\colhead{Object} & \colhead{Date} & \colhead{$v_\mathrm{max}$\tablenotemark{a}} & \colhead{$\sigma_v$\tablenotemark{b}} & \colhead{Instrument} \\
\colhead{} & \colhead{} & \colhead{(km s$^{-1}$)} & \colhead{(km s$^{-1}$)} & \colhead{}}
\startdata
Gaia17bpi & 2022-07-31 & 206 & 60 & NIRSPEC \\
  & 2023-10-31 & 232 & 54 & NIRSPEC \\
\hline
HBC 722 & 2016-11-19 & 123 & 11 & IGRINS \\
 & 2022-07-31 & 102 & 7 & NIRSPEC \\
 & 2022-12-03 & 102 & 15 & NIRSPEC \\
 & 2023-10-30 & 98 & 19 & NIRSPEC \\
\hline
Parsamian 21 & 2018-10-25 & 126 & 26 & IGRINS \\
 & 2022-07-31 & 113 & 14 & NIRSPEC \\
 & 2023-11-01 & 113 & 24 & NIRSPEC \\
 \hline
V733 Cep & 2022-07-31 & 80 & 8 & NIRSPEC \\
\hline
V1057 Cyg\tablenotemark{c} & 2022-07-31 & 4 & 40 & NIRSPEC \\
 & 2022-12-03 & 4 & 56 & NIRSPEC \\
\hline
V1515 Cyg & 2022-07-31 & 4 & 16 & NIRSPEC \\
\hline
V1735 Cyg & 2022-07-31 & 24 & 12 & NIRSPEC \\
\hline
Gaia18dvy & 2022-08-01 & 52 & 8 & NIRSPEC \\
\hline
PGIR20dci & 2022-08-01 & 32 & 24 & NIRSPEC \\
  & 2023-10-30 & 40 & 24 & NIRSPEC \\
\hline
FU Ori  & 2019-01-20 & 121 & 10 & IGRINS \\
 & 2022-12-03 & 116 & 14 & NIRSPEC \\
  & 2023-11-01 & 116 & 22 & NIRSPEC \\
\hline
RNO 1B & 2022-12-03 & 123 & 32 & NIRSPEC \\
\hline
RNO 1C & 2022-12-03 & 87 & 11 & NIRSPEC \\
\hline
RNO 54 & 2022-12-03 & 95 & 15 & NIRSPEC \\
\hline
V900 Mon & 2022-12-04 & 109 & 18 & NIRSPEC \\
\hline
V960 Mon & 2016-11-18 & 72 & 11 & IGRINS \\
 & 2022-12-04 & 64 & 8 & NIRSPEC \\
 \hline
L1551 IRS 5 & 2016-11-26 & 45 & 7 & IGRINS \\
 & 2023-10-30 & 28 & 24 & NIRSPEC \\
\hline
PP13S & 2023-10-30 & 8 & 32 & NIRSPEC \\
\hline
V2775 Ori & 2019-01-26 & 5 & 14 & IGRINS \\
 & 2023-10-30 & 4 & 32 & NIRSPEC \\
 \hline
BBW 76 & 2023-10-31 & 56 & 52 & NIRSPEC \\
\hline
V883 Ori & 2017-12-26 & 145 & 14 & IGRINS \\
 & 2023-10-31 & 139 & 31 & NIRSPEC \\
 \hline
V582 Aur & 2023-11-01 & 153 & 36 & NIRSPEC \\
\hline
ESO H$\alpha$-148 & 2020-11-14 & 28 & 5 & IGRINS \\
\hline
RNO 91 & 2018-06-03 & 99 & 20 & IGRINS \\
\hline
SPICY15470 & 2023-02-19 & 28 & 8 & IGRINS \\
\hline
SPICY35235 & 2023-03-15 & 98 & 25 & IGRINS \\
\hline
SPICY57130 & 2023-03-16 & 31 & 15 & IGRINS \\
\hline
SPICY68600 & 2023-03-17 & 34 & 11 & IGRINS \\
\hline
SPICY68696 & 2023-03-17 & 20 & 14 & IGRINS \\
\hline
WRAY 15-488\tablenotemark{d} & 2020-11-14 & 29 & 9 & IGRINS \\
\enddata
\tablenotetext{a}{$v_\mathrm{max} = \sqrt{GM_*/R_\mathrm{inner}} \sin i$}
\tablenotetext{b}{$\sigma_v$ gives the uncertainty $v_\mathrm{max}$ estimated by the range of $v_\mathrm{max}$ value at which the reduced $\chi^2$ differs by 1.}
\tablenotetext{c}{While V1057 Cyg is included in this table for completeness, the CCFs are likely tracing the extreme wind absorption in this source, rather than disk broadening.}
\tablenotetext{d}{This source remains a candidate FUor, but we include it in this sample on the basis of the presence of the usual characteristic absorption features in its spectrum and its double-peaked absorption lines.}
\end{deluxetable}

\section{Assessing the impact of limited sample size on our ability to probe the mass distribution} \label{app:SampSize}

\begin{figure}[!htb]
    \centering
    \includegraphics[width=0.48\linewidth]{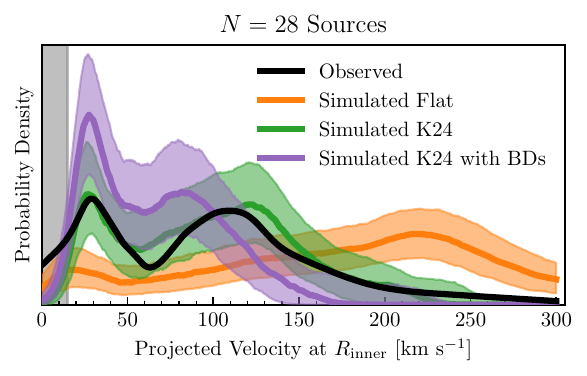}
    \includegraphics[width=0.48\linewidth]{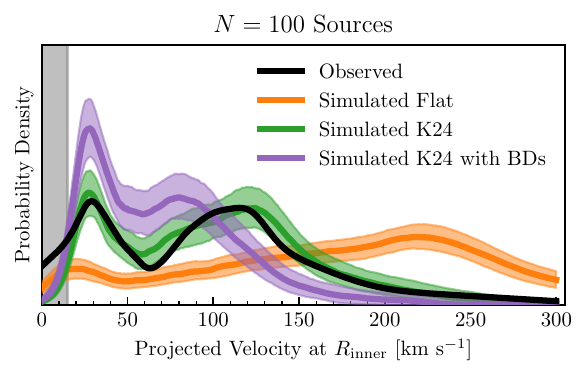}
    \caption{The GSHs for each simulation set with Monte-Carlo-derived error bars assuming a sample of 28 sources (left) and 100 sources (right). See Appendix \ref{app:SampSize} for details. }
    \label{fig:GSH_ebars}
\end{figure}

The sample of FUors for which we have been able to make $v_\mathrm{max}$ measurements is relatively limited due to the lower declination and flux bounds on the NIRSPEC data and the ad hoc nature of the IGRINS sample. In order to quantify the effect of the small, 28 object sample, we adopted a Monte Carlo approach. From each of our sets of simulations, we took 1000 draws of 28 samples each and computed GSHs for each of these draws. Figure \ref{fig:GSH_ebars} shows the median and 16th and 84th percentile GSHs for each simulation, again compared with the observed GSH. We omit the ``puffy'' simulations (see Section \ref{sec:simDist}) from this plot for simplicity. 

To determine, statistically, whether we are able to distinguish the flat IMF from the K24 IMF, we compute the 2-sample Kolmogorov-Smirnoff test for each draw of 28 and our measurements. From the test statistic we compute a $p$ value representing the probability that two samples are drawn from the same distribution. The median $p$ value of each comparison is presented in Table \ref{tab:KSTests}. The desired threshold for determining with 95 \% confidence that one distribution of $v_\mathrm{max}$ values is distinct from the other, we require $p < 0.05$. For a sample size of 28, there is only one set of simulations we can confidently state would be distinguishable from the observed data, and that is the Flat IMF, given its $p = 0.026$. The K24 IMF including BDs also has a relatively low $p$ value but does not reach our threshold of $p < 0.05$. 
%\lahcomm{i learned 0.001 was required to rule out similarity.....remember, this is not a probability, so your 0.05 does not equate to e.g. 95\% confidence, if that is how your chose it.}

Besides probing the main shape of the mass function of FUors, the dependence of $v_\mathrm{max}$ on both $M_*$ and on $R_\mathrm{inner} \sim R_*$ should enable us to constrain the mass-radius relationship of FUors. If so, we might be able to determine whether FUors are systematically puffed-up relative to non-outbursting YSOs due to their large accretion rates, which has been long suspected \citep{hartmann_fu_1996}. The comparison between the sets of draws from two simulations enables us to understand the impact of the small sample by establishing whether, in theory, we should be able to recover the difference between distinct adopted mass-radius relations. However, even in the simulations we cannot distinguish the K24 IMF with BDs from that without. We do, nevertheless, recover a similar $p$ value of 0.026 for theoretically distinguishing the K24 IMF from the Flat IMF. 

Looking to the future, we performed the same set of tests assuming a sample size of 100 FUors. This is only 50\% larger than the estimated currently known population of bonafide FUors and FUor-like objects \citep{CCP_OYCAT_2025JKAS}. Given the increasing rate of discovery of FUors, it is likely that we will reach this sample size in the next $5-10$ years. The results of the GSHs computed from the 100-sample draws shown in the lower panel of Figure \ref{fig:GSH_ebars}. 

The $16 \%-84\%$ envelopes in these GSHs are much narrower than in the 28 sample case, enabling us to more clearly distinguish the different simulated populations. Repeating the KS tests for the 100 sample cases, we find that we would be able to very confidently distinguish the K24 IMF with versus without BDs and the flat IMF. We would still not be able to identify any potential radius inflation at the population level, however.  

\begin{deluxetable}{cccc}
    \tablecaption{The results of the KS test for different combinations of data and draws from our simulations. \label{tab:KSTests}
    }
    \tablewidth{0pt}
    \tablehead{\colhead{Sample 1} & \colhead{Sample 2} & \colhead{Sample Size} & \colhead{Median $p$}}
    \startdata
    & \multicolumn{2}{c}{Data versus Models} & \\
    \hline
    Data & K24 & 28 & 0.549 \\
    Data & Flat & 28 & 0.026   \\  
%    \hline
    Data & K24 Puffy & 28 & 0.549 \\
%    \hline
    Data & K24 with BDs & 28 & 0.222 \\    
%    Data & K24 & 100 & 0.4723 \\
%    Data & Flat & 100 & 0.0023   \\  
%    \hline
%    Data & K24 Puffy & 100 & 0.5638 \\
%    \hline
%    Data & K24 with BDs & 100 & 0.0189 \\
    \hline
    & \multicolumn{2}{c}{Models versus Models} & \\
    \hline
    K24 & Flat & 28 & 0.026 \\
%    \hline 
    K24 & K24 with BDs & 28 & 0.222 \\
%    \hline
    K24 & K24 Puffy & 28 & 0.549 \\
    K24 & Flat & 100 & $3.21 \times 10^{-5}$ \\
%    \hline 
    K24 & K24 with BDs & 100 & 0.006     \\
%    \hline
    K24 & K24 Puffy & 100 & 0.368 \\
    \hline
    \enddata
\end{deluxetable}

\end{document}